\documentclass[aps,pra,preprint,groupedaddress,showkeys]{revtex4}

\begin{document}

\title{Energy lowering of current-carrying single-particle states in 
open-shell atoms due to an exchange-correlation vector potential}
\author{E. Orestes}
\author{A. B. F. da Silva}
\affiliation{
Instituto de Qu\'{\i}mica de S\~ao Carlos\\
Universidade de S\~ao Paulo\\
Caixa Postal 780\\
13560-970 S\~ao Carlos SP, Brazil}
\author{K. Capelle}
\affiliation{
Instituto de F\'{\i}sica de S\~ao Carlos\\
Universidade de S\~ao Paulo\\
Caixa Postal 369\\
13560-970 S\~ao Carlos SP, Brazil}
\date{\today}

\begin{abstract}
Current-density-functional theory is used to perturbatively calculate 
single-particle energies of open-shell atoms prepared in a current-carrying
state. We focus on the highest occupied such energy, because its negative
is, in principle, the exact ionization energy. A variety of different density 
functionals and calculational schemes are compared with each other and 
experiment. When the atom is prepared in a current-carrying state, a 
current-dependent exchange-correlation functional is found to slightly lower 
the single-particle energy of the current-carrying orbital, as compared to a 
calculation using standard (current independent) density functionals for
the same system. The current-dependent terms in the exchange-correlation
functional thus provide additional stabilization of the current-carrying state.
\end{abstract}


\maketitle

\newcommand{\be}{\begin{equation}}
\newcommand{\ee}{\end{equation}}
\newcommand{\bea}{\begin{eqnarray}}
\newcommand{\eea}{\end{eqnarray}}
\newcommand{\bi}{\bibitem}

\renewcommand{\r}{({\bf r})}
\newcommand{\rp}{({\bf r'})}

\newcommand{\ua}{\uparrow}
\newcommand{\da}{\downarrow}
\newcommand{\la}{\langle}
\newcommand{\ra}{\rangle}
\newcommand{\dg}{\dagger}

\section{\label{intro}Introduction}

Density-functional theory (DFT) \cite{kohn,dftbook,parryang,kohnbeckeparr}
is widely recognized as a powerful reformulation of the many-electron problem
in terms of the charge-density distribution $n\r$. In practise the most
common version of DFT is spin-DFT (SDFT), which employs spin-resolved charge
densities $n_\ua\r$ and $n_\da\r$. While SDFT satisfactorily accounts for
magnetic effects associated with the electron spin, it does not explicitly 
deal with magnetic effects associated with the current density. In the presence 
of symmetry-broken states with orbital currents (or in strong external fields), 
a useful alternative to SDFT is provided by current-DFT (CDFT) 
\cite{vr1,vr2,vr3}, which allows explicit calculation of the orbital currents 
and their effect on, e.g., total and single-particle energies.
In CDFT the exchange-correlation ($xc$) energy $E_{xc}$ depends on 
$n\r$ and the {\it current density} 
\be
{\bf j}_p\r=\frac{1}{2i}\sum_k
\left[\psi_k^*\r\nabla\psi_k\r-(\nabla\psi_k^*\r)\psi_k\r\right]
\label{jcdft}
\ee
(atomic units are used in all equations). The derivative of the functional
$E_{xc}[n,{\bf j}_p]$,
\be
{\bf A}_{xc}\r =
\frac{1}{c} \frac{\delta E_{xc}[n,{\bf j}_p]}{\delta {\bf j}_p\r},
\ee
gives rise to an $xc$
vector potential ${\bf A}_{xc}$ in the Kohn-Sham equations, in addition to
the usual scalar $xc$ potential $v_{xc}\r=\delta E_{xc}/\delta n\r$ 
\cite{vr1,vr2,vr3}. Within
(S)DFT ${\bf A}_{xc}\equiv 0$. Consequences of a nonzero ${\bf A}_{xc}$
are little explored in quantum chemistry.

Both SDFT and CDFT share the fundamental property that the negative of the
highest occupied single-particle eigenvalue equals the exact first 
ionization energy $I$ \cite{footnote1}. It is well known that in actual 
(approximate) SDFT
calculations the highest occupied eigenvalue provides only an unsatisfactory
approximation to $I$ if the local-density approximation (LDA) or any of
the common generalized-gradient approximations (GGA's) are used. The main
reason for this failure is incomplete self-interaction correction (SIC). We 
have recently explored numerically the question if the application of CDFT
corrections on top of a converged SDFT LDA calculation can mitigate this 
problem, or reduce the remaining difference to experiment after a SIC has
been applied \cite{pcdft}. 

This investigation was partially motivated by the observation that in order 
to prepare a current-carrying single-particle state in an atom one must 
selectively occupy certain $m$-substates, and leave others unoccupied. 
In the absence of orbital currents and external fields all
single-particle states differing only in the occupation of $m$-substates
are degenerate, but in the presence of currents this degeneracy is broken,
and it becomes a legitimate question to ask if one of the resulting
single-particle energies is a better approximation to the experimental
$-I$ than the one obtained by spherical averaging (i.e., restoring the 
original degeneracy). This becomes an interesting question in particular 
for open-shell systems, where orbital currents can flow.
The numerical results obtained in Ref.~\cite{pcdft} did not allow to give 
a conclusive answer since, although the proposed calculation turned out to be
numerically feasible, no systematic improvement of ionization energies with 
respect to SDFT was achieved. While this was discussed in some detail in
Ref.~\cite{pcdft}, another interesting aspect of those data was mentioned 
there only in passing, namely that the current-dependent correction tends to
lower the single-particle energy, as compared to an (S)DFT calculation 
for the same state.

This lowering suggests that inclusion of the current-dependent part of the 
full exchange-correlation functional --- and hence of ${\bf A}_{xc}$ ---
provides extra stabilization of the 
current-carrying state, {\it relative to a current-independent calculation
for the same state}. 
In the present paper we continue this investigation of current-carrying
single-particle states, and show that the lowering of the single-particle
energy of the current-carrying orbital due to $xc$ effects is
robust against a variety of technical and conceptual changes in the 
calculational procedure.

\section{\label{methods}Methods} 

In general, the ionization energy of a many-body system is defined as            
$I=E(N-1)-E(N)$, where $E(M)$ is the many-body ground-state energy of 
the $M$-electron system. Similarly, the current flowing in a many-body 
system in the absence of externally applied magnetic fields is the 
expectation value of
\be
\hat{\bf j}_p\r =
\frac{1}{2i}\sum_i^N
\nabla_i\delta({\bf r}-{\bf r}_i)+\delta({\bf r}-{\bf r}_i)\nabla_i,
\ee
taken with the many-body wave function. Within DFT, SDFT and CDFT the
ionization energy can also be calculated from the negative of the highest 
occupied single-particle eigenvalue \cite{footnote1}, and within CDFT 
(but not DFT or SDFT) the current can also be calculated from 
Eq.~(\ref{jcdft}), which is obtained by taking the expectation value of 
$\hat{\bf j}_p\r$ with the Kohn-Sham Slater determinant
\cite{vr1,vr2,vr3}. Note that these are exact properties,
which hold regardless of the fact that the quantum numbers $(l,m)$ used to
label the noninteracting atomic states are not the same used for the
interacting states, $^{2S+1}L_J$.
 
Consequently, it is not necessary to construct current-carrying many-body 
states to calculate the ionization energy of the current-carrying system. 
The (at least $m$-fold degenerate) highest eigenvalue of the unperturbed 
noninteracting Kohn-Sham system already yields the ionization energy in the 
absence of currents. A current-carrying state is then described by selectively 
occupying single-body substates with definite value of $m$, obtained by
multiplying a numerical radial function with the appropriate spherical 
harmonic $Y_m^l(\theta,\Phi)$. Within SDFT these current-carrying states 
are still degenerate, but CDFT is capable of picking up the energetic 
contribution of the current to the highest occupied single-particle state,
and will thus in general make a different prediction for the ionization 
energy. The original motivation for this work was to see if the resulting 
ionization energies of current-carrying single-particle
states were better approximations to the experimental ones,
in particular for atoms whose many-body configuration has $L\neq 0$ or 
$J\neq 0$. This expectation was not bourne out by the numbers in 
Ref.~\cite{pcdft}. The present paper is concerned with the question whether 
this is due to computational approximations made in that work, or a real
physical phenomenon.

The computational approach of Ref.~\cite{pcdft} was to apply CDFT not
self-consistently, but perturbatively, following a converged (S)DFT
calculation. This strategy, denoted perturbative CDFT (pCDFT) was proposed
by one of us in Ref.~\cite{jpert}, and leads to a major simplification of 
the CDFT equations. While similar in spirit to post-LDA applications of 
GGA \cite{b88}, or post-GGA applications of Meta-GGA \cite{becke,metagga},
pCDFT is not simply obtained by substituting SDFT orbitals in CDFT
expressions (as in `post' methods), but amounts to low-order perturbation
theory with respect to ${\bf A}_{xc}$, taking as unperturbed system the
Kohn-Sham equations of (S)DFT. In this sense there is some conceptual 
similarity between pCDFT and M\o ller-Plesset perturbation theory, and also 
between pCDFT and the SDFT derivation of the Stoner approximation to the
theory of itinerant ferromagnetism \cite{gunnarsson}.
Since the basic equations of CDFT and pCDFT have been described in detail
in Refs.~\cite{vr1,vr2,vr3} and \cite{pcdft,jpert}, respectively, we 
refer the interested reader to these papers, and here focus directly on
numerical results and their interpretation.
For later reference we record, however, the explicit expression for the
perturbative shift of the $k$'th single-particle eigenvalue due to the 
presence of the current in the orbital $\psi_k\r$ \cite{jpert},
\be
\delta\epsilon_k=
\frac{1}{c}  \int d^3r \, {\bf j}_{p,k}\r \cdot
{\bf A}_{xc}[n,{\bf j}_{p}]\r,
\label{deltae}
\ee
where 
\be
{\bf j}_{p,k}\r = \frac{1}{2i}
\left[\psi_k^*\r\nabla\psi_k\r-(\nabla\psi_k^*\r)\psi_k\r\right].
\ee

Our previous results for the pCDFT corrections to the highest occupied
DFT-LDA eigenvalue for open-shell atoms prepared in a current-carrying
state indicate that, without applying a SIC, CDFT eigenvalues are about as
far from the experimental first ionization energies as SDFT ones, whereas 
after applying a SIC the remaining difference between experiment and LDA-SIC 
is comparable to (but still larger than) the pCDFT corrections \cite{pcdft}. 
We found the pCDFT corrections to be mostly negative, i.e., to stabilize the 
current-carrying state as compared to a (current-insensitive) SDFT calculation 
for the same current-carrying configuration. Since LDA+SIC sometimes 
underestimates and sometimes overestimates the true ionization energy, a 
negative correction cannot consistently improve agreement with experiment. 
However, the fact that the single-particle energy is lowered upon including 
${\bf A}_{xc}$ is interesting in its own right, even if the deviation of 
LDA-SIC from experimental ionization energies is not dominated by 
current-related effects, and it is important to check whether it is robust
against changes in the approximations and computational procedures.

To further investigate these issues we consider, in the present paper, three 
methodological changes with respect to the
calculations of Ref.~\cite{pcdft}. First, we obtain the orbitals and
energies of the unperturbed system (i.e., the DFT Kohn-Sham equations) not 
with LDA but from two common GGAs: B88-LYP \cite{b88,lyp} and PW91 \cite{pw91}.
Either choice should yield an improved description of the input orbitals
needed for perturbatively calculating the effect of ${\bf A}_{xc}$. In a
first step we made only this change, to single out the consequences of
passing from LDA to GGA without changing anything else.

In a second step, we analyse the role played by the orbital susceptibility
$\chi$ in the CDFT LDA, $E_{xc}^{LDA}[n,{\bf j}_p]$ (cf. Eqs.~(8,9) of 
Ref.~\cite{pcdft} or Eq.~(17) of Ref.\cite{vr1} for the explicit form of this
functional). In the spirit of the ordinary LDA, this functional is derived 
from the interacting uniform electron gas \cite{vr1,vr2,vr3}. 
Current-related many-body effects enter via the orbital (Landau)
susceptibility $\chi$, which has been calculated numerically for the
electron gas by Vignale, Rasolt and Geldart \cite{vrg} and parametrized by 
Lee, Colwell and  Handy (LCH) \cite{handy1,handy2,handy3} and in 
Ref.~\cite{pcdft}. 
We observe that of these two parametrizations the five-term 
interpolation \cite{pcdft}
\be
s_{5}(r_s)=1.1038-0.4990 r_s^{1/3}+0.4423 \sqrt{r_s}
\nonumber \\ -0.06696 r_s  +0.0008432 r_s^2,
\label{fiveterm}
\ee
provides a better fit of the 10 data points of Ref.~\cite{vrg}, 
while the LCH expression \cite{handy1,handy2,handy3}
\be
s_{LCH}(r_s)=(1.0 + 0.028 r_s) \exp{(-0.042 r_s)}
\label{lhc}
\ee
has the correct limiting behaviour as $r_s\to 0$ and $r_s\to \infty$, 
and a smoother derivative. In addition to calculations employing consistently 
the LCH or the five-term interpolations, we thus tentatively also implemented 
a hybrid calculation that employs (\ref{fiveterm}) for $s\r$, but (\ref{lhc}) 
for its gradient
\be
\nabla s\r = \frac{d s(r_s)}{d r_s} \frac{d r_s(n)}{dn} \nabla n\r.
\ee
In these expressions $r_s$ is related to the density via $n =3/4 \pi r_s^3$, 
and $s=\chi/\chi_0$, where $\chi_0$ is the (known) orbital susceptibility of 
the noninteracting electron gas.

Third, we observed that either expression for the orbital susceptibility
is derived from the Vignale-Rasolt-Geldart data for the electron gas
\cite{vrg}, which clearly has a susceptibility that is very different 
from that of an open-shell atom. We thus also considered a semi-empirical
expression of the LCH form,
\be
s_{se}(r_s)=(1.0 + b\, r^c_s) \exp{(-0.042 r_s)}, 
\label{se}
\ee
whose parameters $b$ and $c$ were fitted to reproduce the experimental
value of $-I$ for the Carbon atom and then employed unchanged for the 
other atoms.
This semi-empirical calculation can provide a valuable additional piece of 
information: If it turned out that with a semi-empirical susceptibility, 
fitted to one atom only, good ionization energies were obtained also for 
the other atoms this would suggest that the {\it form} of the employed 
functional for ${\bf A}_{xc}$ is in principle correct, but 
handicaped by the input data from the electron gas. 
If, on the other hand, replacing the electron-gas susceptibility by a 
semi-empirical one did not improve the ionization energies for the other 
atoms considered, this would suggest that the functional itself may not be 
adequate for this type of calculation, independently of the origin of the 
susceptibility used in it.

\section{\label{results}Results and Discussion}

Our results are summarized in Table \ref{table1}, which for each atom and
current-carrying (cc) single-particle (sp) state considered lists
the difference $\Delta I$ between the LDA+SIC data of Ref.~\cite{krieger}
and the experimental data \cite{nist}. The other columns contain data
for the negative of the pCDFT shift of the highest occupied eigenvalue,
$-\delta\epsilon$, obtained from Eq.~(\ref{deltae}) with the various 
calculational schemes described above. In every case the unperturbed DFT 
orbitals and energies were obtained using the fully numerical (basis-set 
free) Kohn-Sham code {\it opmks} \cite{opmks}.

In Table \ref{table1} the current-carrying single-particle states are 
characterized succinctly by the values of $m$ of the single-particle orbitals 
occupied in the open shell. As an example of our notation, for the Carbon atom 
the numbers $\{1,0\}$ in the second column mean that of the two electrons in 
the open p-shell one is in a state with $m=1$, the
other in a state with $m=0$. Note that states with $m=0$ do not make a
contribution to either ${\bf j}_p$ or ${\bf A}_{xc}$. States with $m=1$
and with $m=-1$ make a contribution of same magnitude but opposite sign 
to each ${\bf j}_p$ and ${\bf A}_{xc}$. Hence both lead to the same value
for the resulting energy shift, which is determined by the product
${\bf j}_p \cdot {\bf A}_{xc}$. Here we consider only one single-particle
configuration for each open shell. Ref.~\cite{pcdft} contains data for
other choices of occupation.

We now systematically discuss each of the calculational schemes described
above, and the resulting conclusions.
Altough we do not find that ionization energies are systematically improved
by pCDFT, we still present, in Sec.~\ref{ion}, a rather detailed
discussion of the corresponding numbers, since they justify our
conclusion that the difference between experiment and LDA+SIC is most
likely {\it not} due to current-related effects.
In Sec.~\ref{lower} we then turn to the possible self-induced 
stabilization of current-carrying states via formation of a nonzero
$xc$ vector potential ${\bf A}_{xc}$.

\subsection{\label{ion}Ionization energies}

In column three of Table~\ref{table1} we reproduce the results obtained 
with LDA for $v_{xc}$ and the LCH expression (\ref{lhc}) for $s=\chi/\chi_0$. 
These are the same data listed in Table I of Ref.~\cite{pcdft} and are 
repeated here for comparison purposes \cite{footnote3}. 
As pointed out in Ref.~\cite{pcdft}, there is no overall systematic
trend as to when the pCDFT correction improves the LDA+SIC value and
by what margin, although we note that for the first-row elements there 
is a systematic correlation between the atomic number $Z$ and the size 
of $\Delta I$, while for first and second-row elements there is a 
correlation 
between $\delta\epsilon$ and the size of the current. The deviation
between the LDA+SIC+pCDFT values and experiment is, however, in most
cases larger than the pCDFT correction itself, which leads us to believe
that other effects, not related to orbital currents, must play a more 
important role in explaining the remaining differences. Some possible sources
for these are explored in the following.

The next two columns list data obtained using the B88-LYP and PW91 GGAs for
$v_{xc}$. As anticipated in 
Ref.~\cite{pcdft}, the resulting changes are very small. This shows that
for the remaining calculations we can use either functional, without 
significant changes in the final numbers. It also implies that an improved
treatment of the unperturbed system does not improve agreement with 
experiment, suggesting, as expected, that the critical
ingredient in the calculation is the current-dependent part of the full 
$xc$ functional, not the charge-density dependent one.

The following column contains results obtained with the five-term 
expression (\ref{fiveterm}) for the susceptibility ratio $s=\chi/\chi_0$ 
in the current-dependent part of the functional.
These values replace the ones listed in Table I of Ref.~\cite{pcdft}, which
suffered from a numerical error. This correction does lead to somewhat
different numbers, but does not affect any of the conclusions drawn in
Ref.~\cite{pcdft}. In particular, it remains still true that the numbers are
more strongly dependent on the choice made for the orbital susceptibility 
than on the one for the charge-dependent part of the functional (LDA/GGA). 
The next column contains data from the hybrid implementation
using the LCH expression (\ref{lhc}) for determining $\nabla s\r$ and the 
five-term expression (\ref{fiveterm}) for $s\r$. The results are on average
closer to those obtained with only the LHC expression than to those obtained
with only the five-term interpolation.

Finally, we list data obtained with the semi-empirical
expression (\ref{se}). In this case the value for the $m_1=1$ and $m_2=0$
single-particle (Kohn-Sham) configuration of the Carbon atom was used to 
determine the parameters $b$ and $c$, while the other values were obtained
holding these parameters fixed. The optimal values $b=0.161$ and $c=0.689$
exactly reproduce the first ionization energy of Carbon.
This procedure allows us to disentangle the
form of the functional from the electron-gas origin of the original expression
for the susceptibility. Clearly, some improvement for the other atoms 
is obtained in this way, in particular as regards the sign of the corrections, 
but the improvement is still not fully satisfactory or consistently 
obtained for all considered systems. This observation 
suggests that future work should go into deducing CDFT functionals
adequate for finite systems, instead of improving on the electron-gas data.

Our conclusion from all these calculations is that currents in open shells
do not make a decisive contribution to atomic ionization energies.

\subsection{\label{lower}Possible self-stabilization of current-carrying states}

We now focus on the CDFT shifts $\delta\epsilon$ themselves, independently
of the question how they change ionization energies. For this purpose we 
have to disregard the second-to-last (semi-empirical) column of 
Table~\ref{table1}, because the values of $\delta\epsilon$ listed there 
were forced to have the sign and size required to obtain perfect agreement 
with experimental ionization energies 
for the Carbon atom, and can thus not be used to discuss sign and size 
of the {\it calculated} correction.

A notable feature of the non-empirical data collected in Table \ref{table1} 
is that, with exception of a few cases using the five-term interpolation 
for the derivative of $\chi$, 
the pCDFT correction $\delta\epsilon$ is negative throughout.
(Data obtained with the five-term interpolation for the derivative of the
orbital susceptibility may be less reliable because of the wrong $r_s\to 0$ 
and $r_s\to \infty$ limits and the polynomial fit involved.) To appreciate 
that this lowering of the single-particle
eigenvalue of the current-carrying state is not entirely
trivial, consider the following two observations:
(i) Both ${\bf j}_p\r$ and ${\bf A}_{xc}\r$ individually have positive and
negative values for some ${\bf r}$, it is only the integral over their
product, Eq.~(\ref{deltae}), which is negative. 
(ii) ${\bf A}_{xc}$ itself is nonzero 
only due to the presence of the current, so that the energy lowering is not 
simply due to the accomodation of a current in the system in response to some 
external field.

On the single-particle level, inclusion of the $xc$ vector potential thus 
provides additional stabilization of the current-carrying state, as compared 
to ordinary DFT or SDFT calculations, which are
insensitive to the current. While this stabilization is apparently not a 
decisive factor in determining ionization energies, it may have significant 
consequences in other situations, since it implies that a proper assessment
of the energetics of processes involving electron flow 
should consider the $xc$ effects associated with charge currents,
and not only those associated with the charge density. In the present case
the pCDFT shifts are relatively small, but within the accuracy of modern
density functionals. A systematic exploration of current-related many-body 
effects could hold some surprises, in particular for nearly degenerate or
symmetry-broken states, or in the presence of strong external fields. 

We stress that this energy lowering is obtained for a fixed current-carrying
state, comparing a calculation that is insensitive to the current (the LDA 
of SDFT) with a calculation that explicitly accounts for current-dependent 
correlations (the LDA of CDFT, implemented within pCDFT). A related result 
was recently reported in Ref.~\cite{becke} --- employing another
current-dependent density functional, not constructed within the framework 
of CDFT --- where the total energies of some current-dependent 
states were found to be lower in a current-dependent calculation than 
in a current-independent one.

In this context it is also interesting to recall the suggestion by Rasolt and
Perrot \cite{rasper} that the {\it ground state} of a strongly inhomogeneous 
many-body system can develop spontaneous self-induced currents.
This result was obtained using the same formal framework as here (CDFT),
but with a quite different choice of the density functional (optimized
for two-dimensional systems), and by performing a direct minimization
of the total energy (thus avoiding any self-consistent calculations).
Clearly a spontaneously current-carrying ground state is an extreme example
of self-stabilization, in which the energy lowering due to current-dependent
correlations does not only reduce the energy below the one of a 
current-independent calculation for the same current-carrying state, but even
below the one of the not current-carrying state. It remains to be
explored whether our above result, obtained for single-particle
energies, can be related to this type of novel many-body ground state.

\section{\label{summary} Summary}

Concerning the calculation of ionization energies, we find that neither the 
description of the charge-dependent part of the functional (LDA or GGA) 
nor the quality of the
interpolation used for the electron-gas susceptibility (LCH or five-term 
interpolations) decisively influence the quality of the final results, whereas
a semi-empirical expression for the susceptibility yields better results 
than expressions based on the electron gas. Although it is thus possible that
current-dependent functionals designed specifically for finite systems will
further improve results, as compared to electron-gas based functionals,
the remaining differences to experiment are sufficiently large to suggest that
they are not due to orbital currents.

While ionization energies thus do not seem to be systematically affected by
current-dependent corrections, the single-particle energy of the 
current-carrying states is. The self-stabilization of these states by means of 
the exchange-correlation vector potential, ${\bf A}_{xc}[n,{\bf j}_p]$, is
found to be robust against a variety of numerical and conceptual changes in 
the computational procedure. This self-stabilization
is completely missed in current-independent calculations, but may be relevant 
for studies of the energetics of processes involving electron flow and 
states with orbital currents in general.\\

{\bf Acknowledgments} We thank E. Engel for providing us with the 
Kohn-Sham code {\it opmks}, which was used for obtaining the (S)DFT 
orbitals and energies within the LDA, B88-LYP and PW91 approximations.
We also thank T. Marcasso for collaboration at an earlier stage of
this project, and L.~N.~Oliveira for useful discussions. Financial support 
by FAPESP and CNPq is gratefully acknowledged.

\newpage

\begin{table*}
\caption{\label{table1} Current-induced changes in the ionization energies
of atoms with open $p$ ($B$ to $Cl$) and $d$ ($Sc,Y$) shells.
Column one: atom.
Column two: selected current-carrying (cc) single-particle (sp) state,
characterized by the occupied $m$-substates in the open shell. 
As in Ref.~\cite{pcdft} we have
normally considered several current-carrying configurations for each atom,
but here we list only one for each, since the numbers for the others do not
affect any of the conclusions.
Column three: Negative of the pCDFT correction obtained with LDA, using the 
LCH expression (\ref{lhc}) for the susceptibility $\chi$.
Column four: Negative of the pCDFT correction obtained with B88-LYP GGA, using 
the LCH $\chi$.
Column five: Negative of the pCDFT correction obtained with PW91 GGA, using the 
LCH $\chi$.
Column six: Negative of the pCDFT correction obtained with LDA using the
five-term (5t) Eq.~(\ref{fiveterm}) for $\chi$.
Column seven: Negative of the pCDFT correction obtained with LDA, using 
Eq.~(\ref{fiveterm}) for $\chi$ itself, but the LCH expression 
(\ref{lhc}) for its gradient.
Column eight: Negative of the pCDFT correction obtained with LDA, using the 
semi-empirical (se) expression (\ref{se}) for $\chi$.
Column nine: deviation of zero-current ionization energy calculated within
LDA-SIC \protect\cite{krieger} from experimental ionization energies
\protect\cite{nist}.
All values are in $eV$.}

\begin{ruledtabular}
\begin{tabular}{r|rrrrrrrr}
& cc sp & $-\delta\epsilon$& $-\delta\epsilon$ & $-\delta\epsilon$ &$-\delta\epsilon$&$-\delta\epsilon$&$-\delta\epsilon$&\\
& state & LCH & LCH    & LCH  & 5t  & 5t/LCH & se & $\Delta I$ \\
&             & LDA & B88LYP & PW91 & LDA & LDA   & LDA & \\
\hline
B  &1& 0.072&0.070&0.071&0.016&0.034&-0.32& -0.018\\
C  &1,0& 0.045&0.044&0.044&-0.012&0.023&-0.34& -0.34\\
N  &1,1,0&0.14&0.14&0.14&-0.085&0.086&-1.3&-0.42\\
O  &1,1,0,0&0.11&0.11&0.11&-0.10&0.39&-1.2&-0.71\\
F  &1,1,0,0,-1&0.023&0.023&0.023&-0.032&0.17&-0.27&-1.2\\
Al &1&0.043&0.043&0.044&0.025&0.027&-0.092&-0.42\\
Si &1,0&0.022&0.021&0.021&0.0061&0.016&-0.093&0.35\\
P  &1,1,0&0.059&0.059&0.059&0.0082&0.046&-0.34&0.42\\
S  &1,1,0,0&0.044&0.044&0.044&-0.0054&0.082&-0.32&-0.050\\
Cl &1,1,0,0,-1&0.0091&0.0091&0.0091&-0.0017&0.020&-0.073&-0.11\\
Sc &1&0.036&0.036&0.036&0.012&0.043&-0.26&-\\
Y  &1&0.035&0.034&0.034&0.015&0.035&-0.19&-
\end{tabular}
\end{ruledtabular}
\end{table*}

\end{document}